\pdfoutput=1
\documentclass[prl,twocolumn,amsmath,amssymb, pra, aps]{revtex4}
\usepackage{graphicx}
\usepackage{mathrsfs}
\usepackage{subfigure}
\DeclareMathAlphabet{\mathpzc}{OT1}{pzc}{m}{it}
\usepackage{color}

\newcommand{\isat}{I_\textrm{sat}}

\begin{document}

\title{Spectral Lineshape Measurements with Shot-Noise Limited Accuracy} 

\author{Gar-Wing Truong}
\email[]{Gar-Wing.Truong@uwa.edu.au}

\author{James D. Anstie}
 \affiliation{Frequency Standards and Metrology Research Group, School of Physics, The University of Western Australia, Perth, Western Australia 6009, Australia}

\author{Eric F. May}
\affiliation{Centre for Energy, School of Mechanical and Chemical Engineering, The University of Western Australia, Perth, Western Australia 6009, Australia}

\author{Thomas M. Stace}
 \affiliation{ARC Centre of Excellence for Engineered Quantum System, University of Queensland, Brisbane, Queensland 4072, Australia}

\author{Andr\'{e} N. Luiten}
 \affiliation{ Frequency   Standards and Metrology Research Group, School of Physics, The University of Western Australia, Perth, Western Australia 6009, Australia}

\date{\today}

\begin{abstract}

Spectroscopy has played the key role in revealing, and thereby understanding, the structure of atoms and molecules. A central drive  in this field is the pursuit of higher precision and accuracy so that ever more subtle effects might be discovered. Here, we report on laser absorption spectroscopy that  operates at the conventional quantum limit imposed by photon shot-noise. Furthermore, we achieve this limit without compromising the accuracy of the measurement.  We demonstrate these properties by recording an absorption profile of cesium vapor at the 2 parts-per-million level. The extremely high signal-to-noise ratio allows us to directly observe the homogeneous lineshape component  of the spectral profile, even while in the presence of  Doppler broadening that is a factor of 100 times wider. We can do this because we can precisely measure the spectral profile at a frequency detuning more than 200 natural linewidths from the line center. We use the power of this tool to demonstrate direct measurements of a low-intensity optically-induced broadening process that is quite distinct from the well-known power broadening phenomenon. 

\end{abstract}


\maketitle

Researchers performing absorption spectroscopy have continually sought to improve the precision with which atomic and molecular lines can be measured. As instrument precision improved, comparisons of the observed spectra to contemporary theory often led to the discovery of subtle structures that were previously unresolvable. Exemplary cases include the observation, and consequential understanding, of the Stark\cite{stark1913englishletter} and Lamb shifts\cite{lamb1947}. Attaining accuracy as well as precision in modern frequency metrology now rests on the detailed removal of these and many similar systematic frequency effects at the level of a few parts in $10^{17}$\cite{oskay2006,takamoto2005}. Analogously, recent attainment of astonishing frequency precision and purity in the modes of an optical frequency comb are offering new possibilities for laboratory-scale searches for variations in the values of fundamental constants\cite{haensch2005}. Here, it is again demonstrated that achieving incredible precision is only half the story; accuracy at the same level is also required, which demands stringent understanding of all systematic factors in the model for the observed spectroscopic lineshape\cite{niering2000}. Progress in precision laser absorption spectroscopy has thus become an enabling technology of key importance in a wide variety of fields, which include frequency metrology\cite{udem2002,knappe2005}, direct optical frequency comb spectroscopy\cite{marian2004,diddams2007}, primary thermometry\cite{djerroud2009,castrillo2009,lemarchand2010,truong2011}, trace gas detection\cite{thorpe2008,zahniser1995}, precision lineshape measurements\cite{borde2009,arroyo1993} and precision lineshape analysis\cite{cygan2012}. 

The conventional limit to precision in laser absorption spectroscopy (LAS) is set by shot-noise in the probing light, though, in practice, it is difficult to achieve this limit as technical noise and instrumental limitations usually intervene. These technical limitations are particularly acute for LAS because it is inherently a \emph{bright field}\cite{hobbs2009} measurement i.e. the signal of interest is a difference between two large signals: one proportional to the incident power, and the second proportional to the transmitted power. If one wishes to build a high-sensitivity and high-accuracy tool for probing optical absorbance then there are two key challenges: first, small signals are easily obscured by technical noise sources such as laser amplitude instability, which are typically much larger than fundamental sources; and second, the resolution of the measurement (the small differential signal) is compromised  because most of the dynamic range of the sensor  is  consumed by the need  to measure the large average value  of the signal. Ingenious differential-detection designs, such as the \emph{noise eater} described in Ref. \cite{hobbs1991}, can be used to reach the shot-noise limit in the presence of amplitude noise that can up to 70\,dB larger through a high degree of common mode rejection. This  type of balanced detector was recently  demonstrated in conjunction with an optical frequency comb to give high-precision broadband measurements at the shot-noise limit\cite{foltynowicz2011}.

Although the precision of such techniques is extremely impressive, it is important to note that this often comes at the expense of accuracy because information about the average power (common mode signal) is either not measured, or because the design of the differential detector does not adequately ensure linearity. The latter problem is typically  the case for auto-balancing photocurrent attenuators\cite{hobbs1991,hobbs1997,nirvana}. This  loss in accuracy is not critical if the ultimate goal is to achieve extraordinary sensitivity; however, in many areas of current interest, accuracy is just as important as precision\cite{daussy2007,casa2008}. In such circumstances the strict adherence of the measured lineshape to the assumed model profile is critical to avoiding systematic errors when line parameters are extracted from regression. To this end, other work has previously sought, and found, significant non-linearity in the absorption depth of alkali metal vapors even in the weak-probe regime ($I<\isat/10$)\cite{siddons,sherlock2009,shin2009}. However, these earlier studies were unable to resolve the subtle changes in homogeneous linewidth that are at the origin of the lineshape changes.

One brute-force approach to overcome this challenge would be to use traditional LAS but digitize the two signals of interest  with an extraordinarily large dynamic range device (a possibility canvased in Ref. \cite{foltynowicz2011}). However, as we demonstrate in this paper, it is possible with a simple rearrangement of existing equipment, to obtain a linear differential LAS technique that allows measurements of optical power ratios at the few parts-per-million level with precision limited only by shot-noise on the light.

We make  use of this high-resolution spectrometer to probe the D1 transition of cesium (Cs) and report  an unexpected source of optically-induced modification to the homogeneous lineshape component in the absorption profile. Whilst it is well established that the homogeneous linewidth can be substantially modified at intensities higher than the \emph{saturation intensity}\cite{siegman,demtroeder1981}, $\isat$, we directly observe that the linewidth has doubled at intensities as low as $\isat/10$. This effect is an unforeseen consequence of optical pumping in the alkali vapours.

Our high-accuracy, high-precision technique can find application in lineshape analysis to extract collisionally-induced perturbation parameters\cite{deVizia2011,cygan2012}, verification of time-dependent optical pumping models\cite{stace2010}, and studies of plasma condition by lineshape analysis\cite{miura2011,xiong2011}


\section{Results}
 \begin{figure}[ht]
 \includegraphics[scale=0.2]{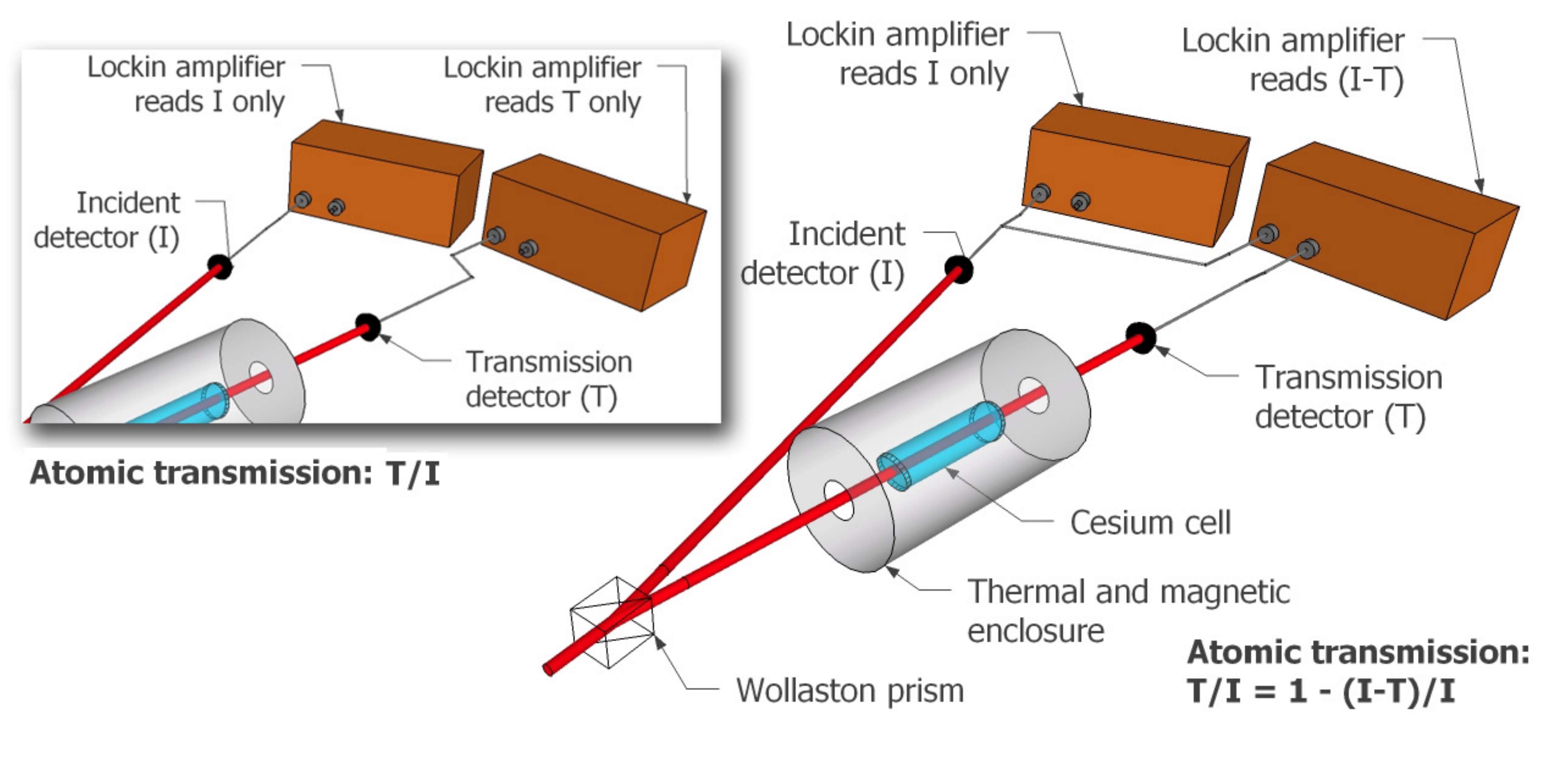}
 \caption{(Color Online) Quasi-differential detection scheme for highly linear, large dynamic range bright field measurements of spectral lines. A frequency-stabilized laser is split at the Wollaston prism. The reference beam is detected on photodiode \emph{I}. The probe beam passes through the Cs cell inside a thermally and magnetically controlled environment, and the transmitted light is detected on photodiode \emph{T}. Inset: Traditional LAS detection technique with sub-optimal resolution limited by instrumental dynamic range. The optical setup is unchanged. \label{fig:simplesetup}}%
 \end{figure}

In conventional LAS detection, the incident and transmitted  light is independently detected (see inset on  Fig. \ref{fig:simplesetup}) to deliver a photocurrent of  $I$ and $T$ respectively\cite{demtroeder1981,lemarchand2010}. Any common-mode laser intensity fluctuations are removed by constructing the ratio $T/I$. This procedure intrinsically compromises the resolution of an absorption measurement as most  of the dynamic range of the photocurrent measurement device is used to deliver solely information on the average photocurrent.   In contrast, our technique is based on effectively extending the dynamic range of the photocurrent measurement by making two different measurements: first, a high-gain measurement of the difference signal $I-T$ which is centred on zero,  and a separate measurement of the average component ($I$). 
%
The ratio $T/I$ can then be reconstructed in software using the identity
\begin{eqnarray}
\frac{T}{I} = 1-\frac{I-T}{I}.
\end{eqnarray}

The increase in dynamic range of this new technique over the traditional method is equal to the inverse of the depth of the absorption signal (i.e. in our case $\sim1/0.05 = 20$). This condition is set by the maximum increase in gain allowable on the detection of $I-T$ over that of $I$ by itself. It is this careful differential approach that has given us a linear, high dynamic-range ratiometric measurement of the optical powers. We have verified that non-linearity of the detection chain contributes on the order of 100\, parts-per-million detection gain variation between 95 and 100\,\% of the typical operating optical powers, (a few microwatts) which we can correct to first order to the level of the uncertainty of our linearity verification (a few parts-per-million) (see Methods).  We expect that the photocurrent ratio ($T/I$) is an accurate measurement of the transmission $P/P_{0}$ to within this limit.

Our measurements are made at probe intensities between $10^{-3} \isat$ to $10^{-1} \isat$ of the saturation intensity, where $\isat = 2.5$\,mW/cm$^2$\cite{steck2003cdl}. At such low probe intensities, the transmitted probe power $P(f)$ can be modeled using the Beer-Lambert law\cite{siegman,demtroeder1981},
\begin{eqnarray}
P(f) = P_0(f)\exp[-\sigma], \label{eqn:BeersVoigt}
\end{eqnarray}
where $P_0(f)$ is the probe power incident on the atoms at each optical frequency $f$ and $\sigma$ is the (frequency dependent) optical depth.  In this same low intensity limit, we can write $\sigma=\alpha V(f-f_0)$, where $\alpha$ is the on-resonance optical depth; $f_0$ is the optical frequency of the transition and $V(f)$ is a Voigt function \cite{demtroeder1981, borde2009} that describes the real-part of the complex susceptibility of the optical resonance for a thermal ensemble of atoms. The Voigt function is a convolution of a Lorentzian component of the lineshape (homogeneous linewidth) and a Gaussian component (inhomogenous Doppler broadening).  The Lorentzian component can be defined as $L(f)= [1+(f/\Gamma)^2]^{-1}$) where $\Gamma$ is the half-width at half maximum (HWHM) bandwidth, while the Gaussian component, $G(f)= \exp(-f/\delta_{w})$ has a width of $\delta_{w} = \sqrt{(2kT)/(mc^2)}$ where $k$ is the Boltzmann constant, $T$ is the temperature of the vapor, $m$ is the atomic mass and $c$ is the speed of light. Broadening effects from transit time and the probing laser  are negligible in this experiment. Power broadening  will only amount to  an apparent 5\% change in  the homogenous linewidth over the probe intensity range\cite{siegman}. We thus expect  the Lorentzian component to be nearly constant with a width set by he natural lifetime of the transition, i.e.\,$\Gamma=2.29$\,MHz. All of our measurements were made at $T=279$\,K so that all of the recorded spectra should also have a fixed Gaussian component with the same characteristic width ($\sim 207$\,MHz).  

 \begin{figure}
 \includegraphics[scale=1]{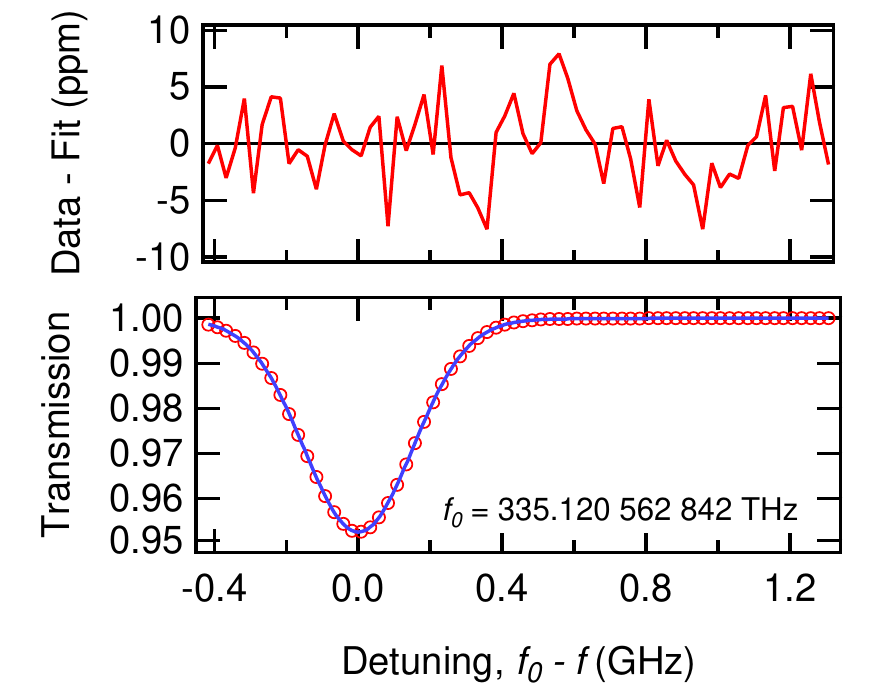}%
 \caption{(Color Online) Lower panel: A typical spectrum of a D1 Cs absorption line. Open circles are the data, and the solid blue is the fit. Upper panel: Residuals to the fit function in Eqn. \ref{eqn:BeersVoigt} and is consistent with a shot-noise limited measurement. \label{fig:spectrum}}%
 \end{figure}


A typical transmission spectrum ($T/I$) for which $I=0.15\isat$ and $T=279.29\pm0.015$\,K is shown in the lower panel of Fig. \ref{fig:spectrum}. A least-squares fit was performed to the spectrum and the residual from a fit to the  Eqn. \ref{eqn:BeersVoigt}  model  is shown in the upper panel of Fig. \ref{fig:spectrum}. The residuals 
are featureless and have a white spectrum with  an root-mean-square (RMS) spread of 2\,ppm \footnote{A small Lamb dip\cite{bennett1962} feature of 60\,ppm amplitude peak and $\sim70$\,MHz wide was visible in the spectra due to weak back-reflections from an uncoated cell window. This feature was included in the fit model to eliminate it from the residuals for clarity.}. This is entirely consistent with the expected photon shot-noise at this probe power (6\, $\mu$W).

\begin{figure}[t]
\centering
\includegraphics[scale=0.7]{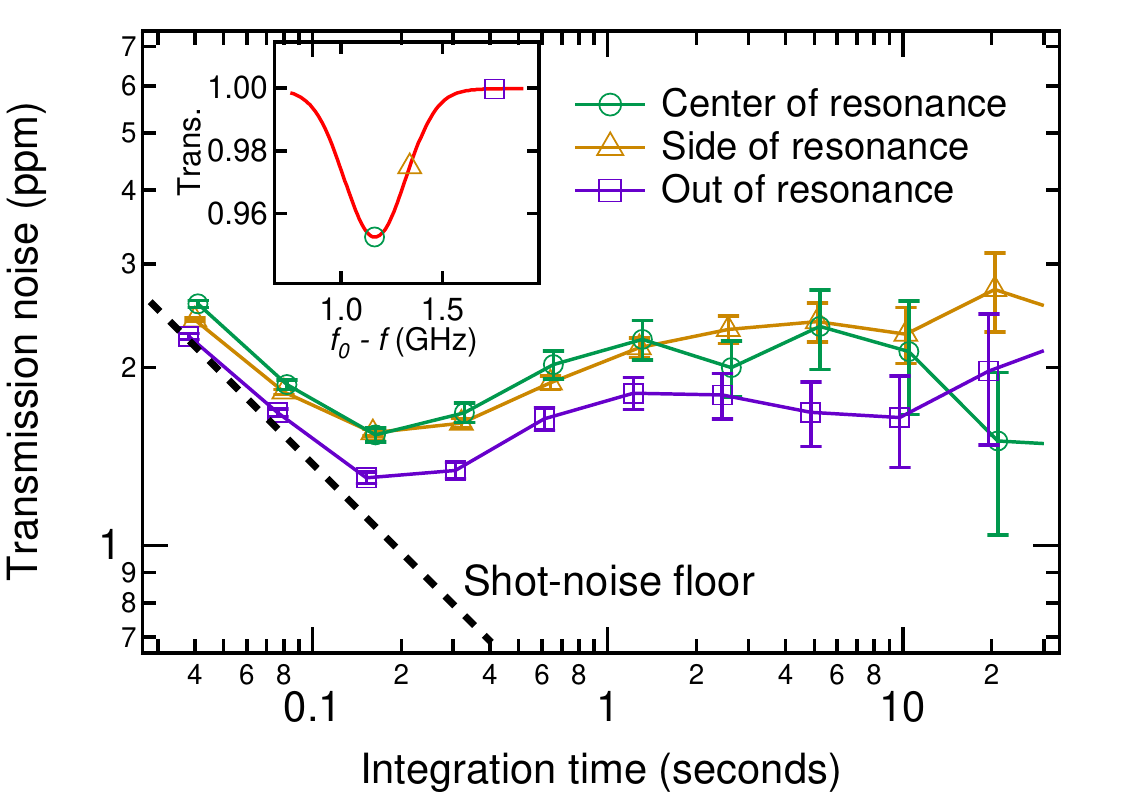}
\includegraphics[scale=0.7]{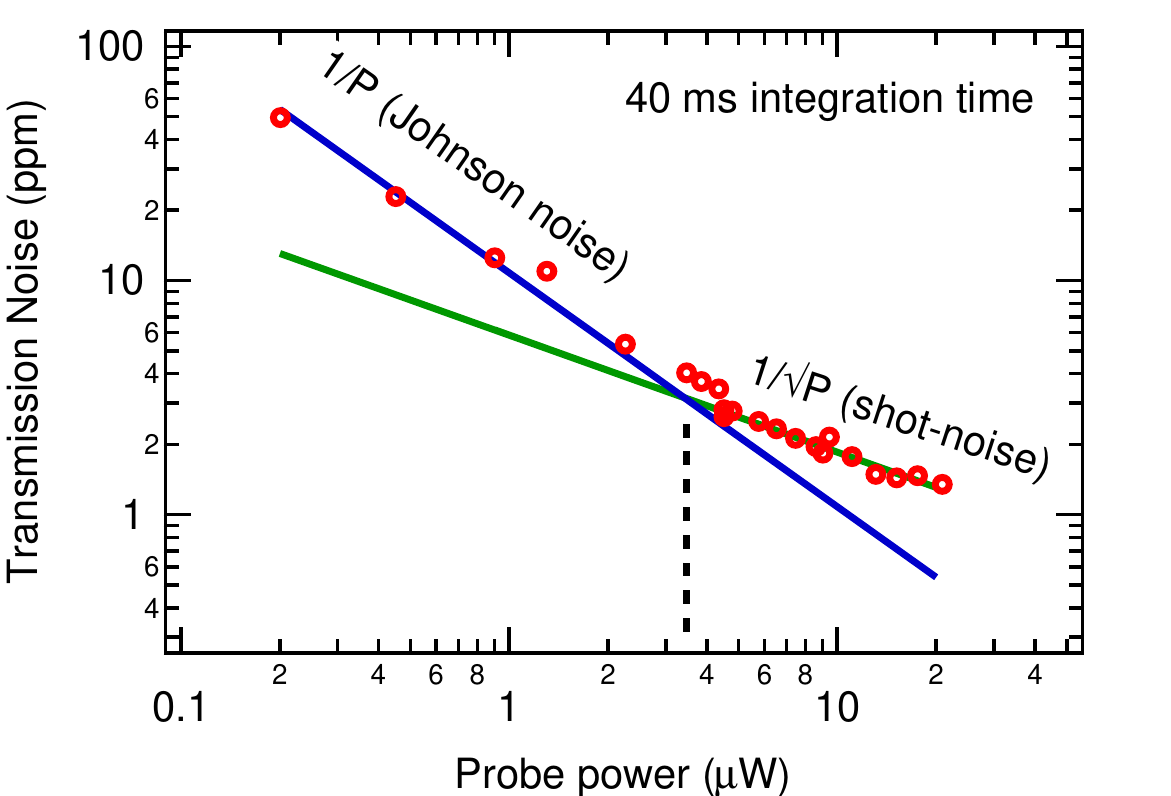}
\caption[Optional caption for list of figures]{Noise in the measurement of optical transmission relative to the off-resonance value of near unity. (Top) The relative noise is white on timescales between 40\,ms to 200\,ms, but becomes dominated by flicker noise at longer times. There is almost no additional noise induced by frequency instability of the probe laser. (Bottom) The relative noise in the measurement of the optical transmission ratio at an integration time of 40\,ms. The measurements become shot-noise limited at probe powers greater than 3\,$\mu$W.}
\label{fig:RINpower}
\end{figure}

To demonstrate explicitly that the measurement noise is associated with shot-noise we monitored the fluctuations in the measurement as  the probe laser was held at three different frequencies corresponding to a) completely out of resonance with the Cs, b) at a point half way down the absorption curve, and c) tuned to the centre of resonance. Each of these measurements should yield similar results if frequency noise of the laser source is negligible (since the transmitted power is similar in all cases). Fig. \ref{fig:RINpower} (top) shows the measured fluctuations in each of these signals over various time scales. We display this in terms of  the Allan Variance which is a standard and stable  measure of fluctuations in an experimentally obtained time series~\cite{allan1966}. The measurement noise is seen to be nearly identical in all cases confirming that the laser frequency noise is negligible at the current measurement precision. The slope of each result between 40\,ms and 0.2\,s of integration time indicates that the fluctuation spectrum  is white. For longer timescales, the measurement is dominated by flicker-noise arising in the electronic components of the measurement system and possible drifts in the alignment of stray etalons\cite{voss1979}. Fig.\ \ref{fig:RINpower}(b) shows the measurement noise at 40\,ms integration time as a function of probe power. Here, we see a distinct change in slope around 3.5$\mu$W incident power  from $1/P$ to $1/\sqrt{P}$ behaviour.  We also display  the calculated Johnson noise of the photodetector readout resistor (blue) and the calculated shot-noise of the light from first principles (green curve). It can be seen that the dependence and magnitude of the noise at low powers is consistent with  Johnson noise, while  the shape and magnitude is consistent with  photon shot-noise at   optical powers higher than 3.5\,$\mu$W.   

The large dynamic range and high SNR  of our approach means that we can directly reveal subtle and important changes in the absorption spectra, which  have not been observed previously.   Fig.\ \ref{fig:logVoigt} shows the frequency-dependent optical depth, $\sigma$, extracted from the data using Eqn.\ \ref{eqn:BeersVoigt}. We display this depth  for  a set of  probe powers ranging from $\sim 10^{-3}- 10^{-1} \isat$.  One observes a strong on-resonance Gaussian profile    associated with Doppler broadening of the vapour.  In addition, the high signal to noise ratio in our experiment enables us to  observe the Lorentzian wings that make up a ``pedestal'' of the Voigt profile at large detunings $\gtrsim100 \Gamma$.  These are easily distinguished: the Gaussian component decays rapidly at high frequency detunings until the  Lorentzian wings become dominant\cite{kitching1993}. 

Fig.\ \ref{fig:logVoigt} shows a surprising, and at first sight quite unexpected, feature: it can be clearly seen  that   magnitude of the Lorentzian component is dependent on the probe intensity.   This is even more  evident on the  inset on this figure where we re-plot the data in terms of its apparent Lorentzian width ($\Gamma$): this increases  from 2\,MHz to 6\,MHz  as the probe intensity increases from $10^{-3}I_\textrm{sat}$ to $10^{-1}I_\textrm{sat}$ whereas we should have expected that this was constant within 5\% over this intensity range\cite{steck2003cdl}.  We have recently shown that this type of  behaviour can arise  out of a complex interplay between optical pumping and atom dynamics   as an atom traverses the probe beam\cite{stace2012}.  This interaction modifies  the underlying Lorentzian lineshape to give an   overall lineshape that is still close to a Voigt function but with a modified Lorentzian width parameter. 

\begin{figure}[t]
\centering
{
\includegraphics[scale=0.7]{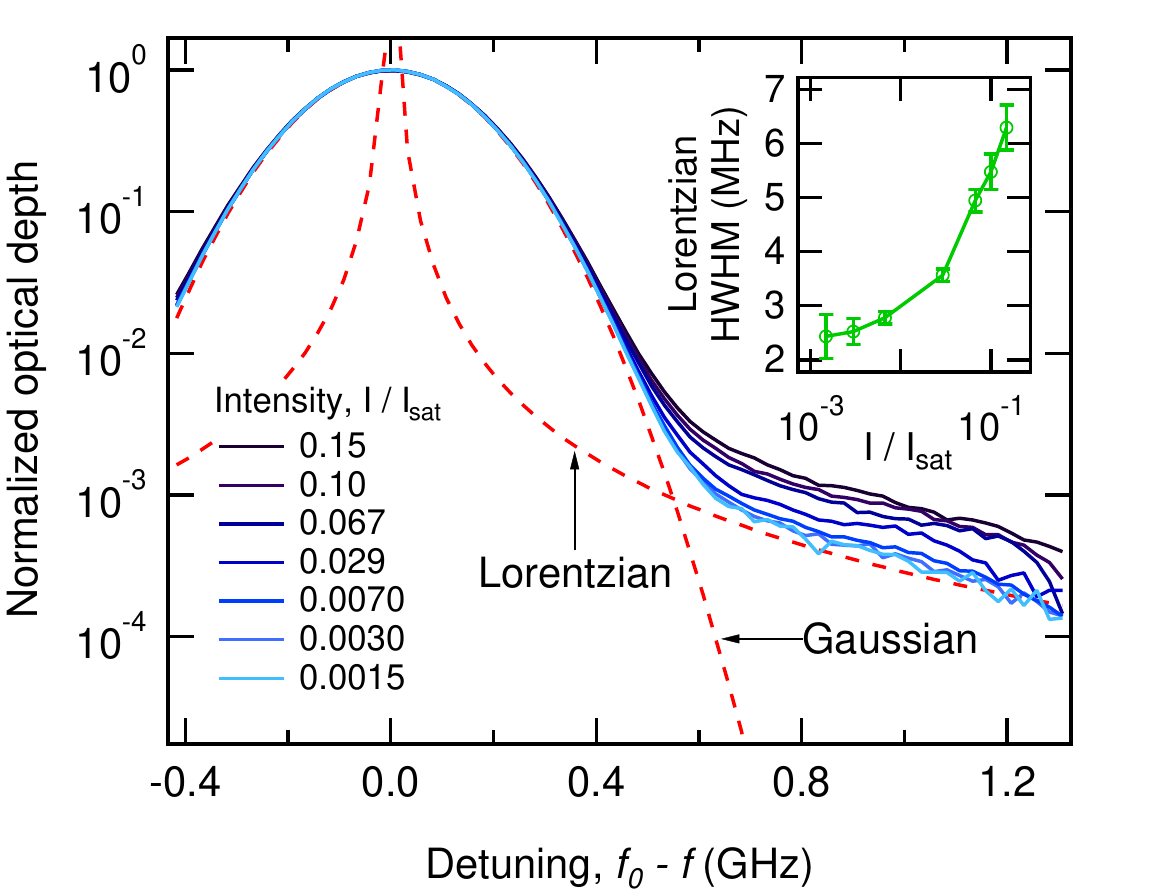}
\label{fig:logVoigt}
}
\caption[Optional caption for list of figures]{(Color Online) Demonstration of the breakdown in the Voigt profile  at low intensities. (top) Optical depth versus detuning for various probe intensities.  The inset shows the fitted Lorentzian width as a function of probe power; the fact that this is not constant over the range of probe intensities is a signature of the breakdown of the Voigt spectral line shape \cite{stace2012}. The reference profile which is expected in the limit of zero probe intensity consists of a Gaussian component of 207.902\,MHz and a Lorentzian component of 2.2875\,MHz.\label{fig:logVoigt}}
\end{figure}

\section{Discussion}
We have demonstrated  a new approach to laser absorption spectroscopy that circumvents the dynamic range limitations of the standard technique. This is achieved by using a  high-precision differential measurement scheme that carefully preserves the measurement linearity while simultaneously obtaining  scale information in order to deliver accuracy. This has permitted the reconstruction of Cs absorption spectra with a SNR limited by shot-noise on the probe source (2\,ppm at 6\,$\mu$W in a 40\,ms bandwidth). By varying the probe power between $1/100$ to $1/10$ of the saturation intensity, we have observed an apparent trebling of the underlying homogeneous linewidth of the Cs D1 transition. This observation shows the power of the new technique as we are making precision measurements of the natural linewidth component of the voigt profile in the wings of the spectral line at frequency detuning of more than 200 natural linewdiths away from the line center. 
This work heralds a potential leap in ultra-high precision absorption spectroscopy of gases where we can now overcome previously overlooked physical phenomena that arise from time-dependent optical pumping. We anticipate these techniques will be critical in future high precision spectroscopy research.

\section{Methods}
 \begin{figure}[ht]
 \includegraphics[scale=0.85]{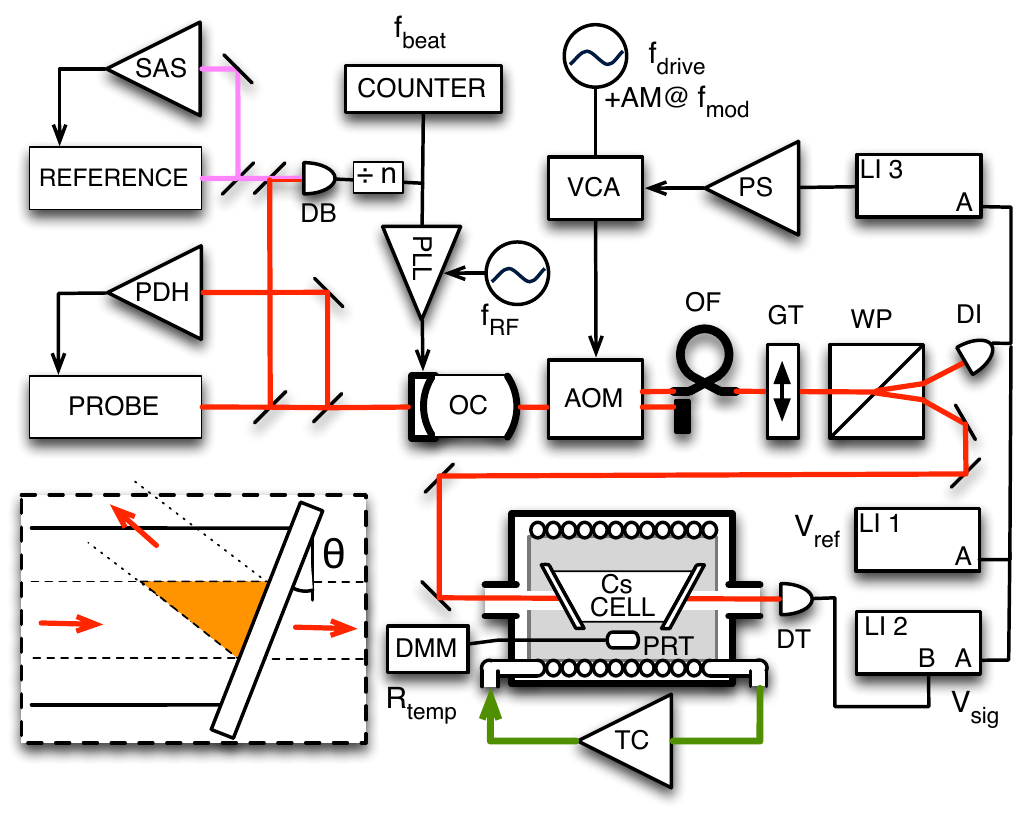}%
 \caption{(Color Online) Optical setup. The inset showns a magnified view of the output window of the Cs cell. There is a small volume overlap, shaded orange, between the incoming probe beam and its weak retroreflection from the angled, but uncoated cell window. $\theta=15^\circ$. \label{fig:setup}}%
 \end{figure}

\subsection{Optical setup}
A detailed schematic of the experimental setup is shown in Fig. \ref{fig:setup}. One extended cavity diode laser (ECDL), labeled \emph{reference}, was frequency-stabilized to a saturated-absorption hyperfine spectroscopic feature of Cs at 894.578\,nm in the D1 manifold, using a standard Pound-Drever-Hall (PDH) laser-locking technique\cite{black2001}. A second ECDL, labeled \emph{probe}, was used to interrogate the cesium transitions at 894.580\,nm, which are 1.17\,GHz detuned from the reference transition. This laser was frequency locked   using the PDH technique to the fundamental transverse mode of an optical cavity (OC) (1\,MHz bandwidth; 18\,GHz free spectral range). A beat-note between the two lasers was formed on a high speed photodiode (DB), whose output was frequency divided by a factor of $n=20480$, and locked to a tunable radio frequency oscillator ($f_{\mathit{RF}}$) using a  frequency-locked loop to stabilize the frequency probe at the 2 kHz level over 1 to 10\,s. By manipulating $f_{\mathit{RF}}$ the probe laser frequency could be tuned to an arbitrary frequency offset from the reference laser.  A frequency counter was used to monitor the (divided) beat-note frequency, $f_{\mathit{beat}}$, which was used to accurately reconstruct the frequency axis for the transition of interest.

Light transmitted by the optical cavity, which was mostly free of   spontaneous emission from the ECDL, was passed through an acousto-optic modulator (AOM). This AOM performed two tasks:  an optical chopper (90\% amplitude modulation at $f_{\mathit{mod}}= 1.57$\,kHz), and as a variable optical attentuator to deliver high-bandwidth power stabilization of the diffracted beam. Th diffracted beam is   shifted in frequency by the AOM's  drive frequency of 175 MHz, $f_{\mathit{drive}}$. The zero-order undiffracted beam was used to compare the probe laser frequency to a stabilised optical frequency comb (not shown). The diffracted beam was coupled into a length of single-mode optical fiber (OF) that  eliminated pointing fluctuations in the light. The resulting spatially and spectroscopically pure probe beam was then polarized and  split using Glan-Taylor (GT) and Wollaston (WP) prisms respectively.  Light from one arm was detected immediately, while light from the other arm was directed through the cesium cell before detection (on DI and DT respectively). 

The detectors  were reverse-biased silicon photodiodes, with matched 47\,k$\mathrm{\Omega}$ load resistors.  The voltage across each load resistor was synchronously detected using a lock-in amplifier (referenced to $f_{\mathit{mod}}$), and recorded on a digital multimeter (DMM).  To circumvent the dynamic range limit imposed by the lock-in amplifier (Stanford Research Systems SR830), which was 10 times worse than the shot-noise limit, a low-noise analog  subtraction stage was used to create a differential signal that was close to zero when the laser was off-resonance. Lock-in amplifier 2 (LI 2) was set to demodulate the differential signal $V_{sig}-V_{ref}$. The advantage of this technique is its ability to measure small transmission deviations from unity without loss of dynamic range. In addition, the incident photodiode (DI) output  was detected by a third  lock-in amplifier to produce a signal for optical power stabilization.  The output of this third lockin was compared to a reference voltage and then fed back to the AOM via a voltage controlled attenuator (VCA) to stabilize the optical power of the probe beam.

The cesium cell under test was encased in a 15\,kg cylindrical copper block. Its temperature was monitored using a using two platinum resistance thermometers (PRTs) calibrated by their manufacturer with an uncertainty of 30\,mK. One PRT was placed at the centre of the block (shown in Fig. \ref{fig:setup}), 5\,mm from the mid-point of the cell; the other was located at the extreme end of the block (not shown). A temperature-controlled chiller sent refrigerated liquid through a copper pipe coiled around the block to maintain a temperature of approximately 276\,K, with stability of 3\,mK over one minute and long term stability of 15\,mK. At this temperature, the cesium vapor in the cell was optically thin. The maximum gradient observed between the PRTs was 30\,mK, which was at the same level as their combined calibration uncertainty. The copper block and cooling coils were enclosed within a mu-metal magnetic shield to minimize Zeeman broadening and mounted on sorbothane pads for passive vibration isolation.

\subsection{Detection Linearity}

The linearity of the entire detection scheme (reverse-biased photodiode and load resistor; lock-in amplifier; and digital multimeter) was determined using a modified, synchronous version of the method discussed by Shin\cite{shin2005}.  Two lasers of equal optical power were combined and passed through an AOM (to produce AM modulation identical to that discussed above) before falling on the photodiode.  Each beam could be independently switched on or off using a mechanical chopper.  A third laser, with optical power 20 times that of each chopped laser, was also passed through the AOM and onto the detector to introduce a signal replicating our operating point.  The two low-power beams were continuously cycled through the four possible states (on:off, on:on, off:on, off:off) and the resulting synchronously detected output voltages were recorded. The figure of merit for this linearity measurement is the relative error, given by: 

\begin{eqnarray}
e = \frac{(V_1 + V_2 ) - V_{12}}{(V_1+V_2)} 
\end{eqnarray}

where $V_1$, $V_2$ and $V_{12}$ are the recorded output voltages for the first beam only, the second beam only and both beams together, respectively.  The relative error can be calculated for each complete cycle with the fourth state (both beams off) used to remove unwanted offsets.  Averaging the results of many such measurements reduces uncertainty in the relative error to arbitrarily small levels.  By measuring the relative error as a function of optical power (or more practically, output voltage) the resulting first-order non-linearity of the detection scheme can be deduced and corrected.



\subsection{Data reduction}
In the least-squares fits of the measured transmission data, a small quadratic power dependence arising from broad and weak etalons formed by the surfaces of each uncoated cell window. This is modeled by allowing the input power in Eqn. \ref{eqn:BeersVoigt} to take the form $P_0(f)= Af^2 + Bf +C$. The parameters $A$, $B$ and $C$ are adjustable fit parameters. To display the underlying Voigt profiles (Fig.\ref{fig:logVoigt}), the parameters $A,B,C$ and $\alpha$ are substituted with their fitted values into 
\begin{eqnarray}
V(f-f_0) = -\ln[P(f)/P_0(f)]/\alpha\,. \label{eqn:isolateVoigt}
\end{eqnarray}

\subsection{Reproducibility of fit parameters}
\begin{figure}[t]
 \includegraphics[scale=0.7]{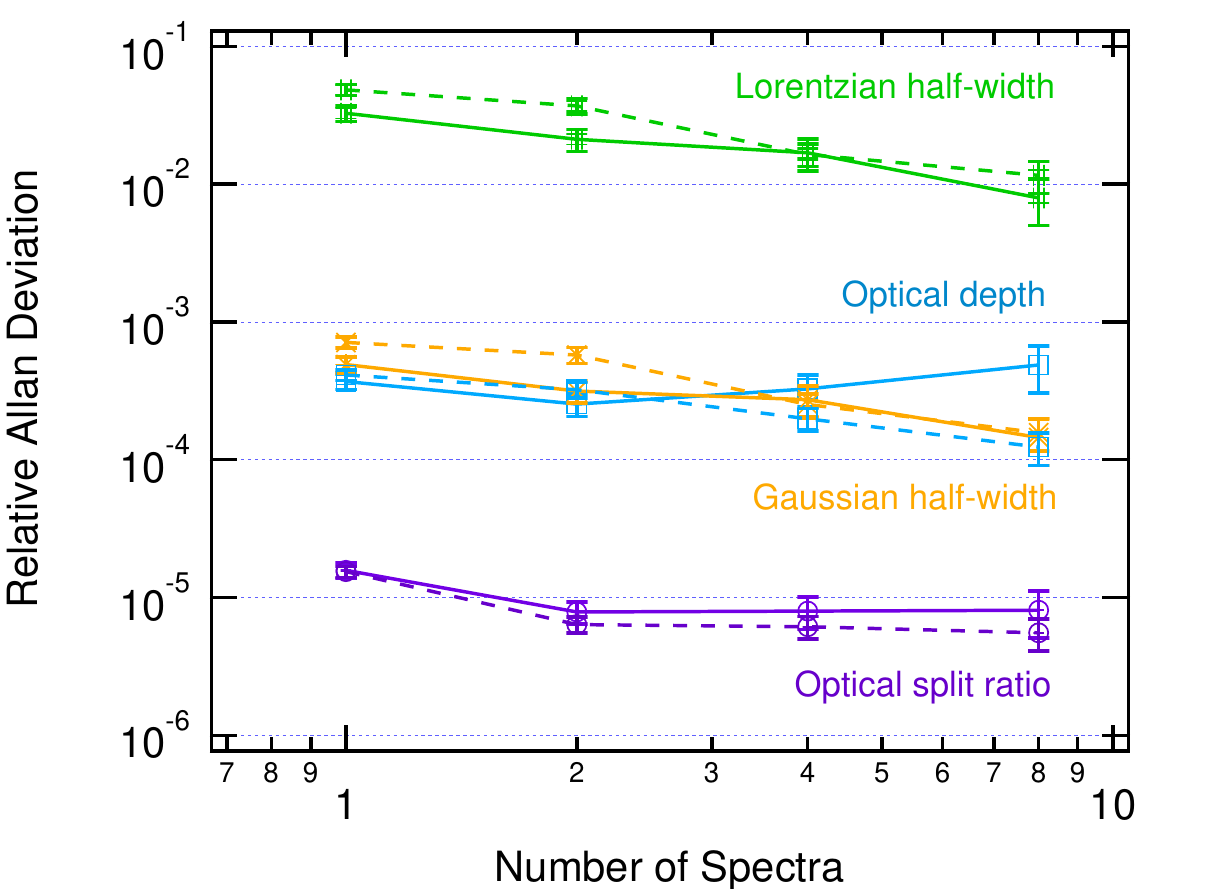}%
 \caption{(Color Online) Relative Allan deviations for selected fit parameters. The solid lines are extracted from measured spectra, whilst the dashed traces are from simulations using noise characteristics consistent with that shown in Fig. \ref{fig:RINpower}a.\label{fig:paramsravs}}%
 \end{figure}

We recorded a set of 60 consecutive scans of the Cs spectrum at a probe power of 6$\mu$W where the measurement is shot-noise limited.  Fig. \ref{fig:paramsravs} displays the relative Allan deviation of a selection of fit parameters in Eqn. \ref{eqn:isolateVoigt} ($C$ the off-resonance ratio, $\alpha$ the absorption depth, $\Gamma$ the Lorentzian half-width and the Gaussian half-width). The dashed traces are the expected parameter noises given the noise characteristics of the transmission noise (Fig. \ref{fig:RINpower}a) and demonstrates that stability of the fitted parameters is consistent with the shot-noise limit in combination with the flicker-noise at longer time scales. The Lorentzian width estimation is 100-fold noisier than the Gaussian because they are highly fit correlated parameters; a unit change in the absolute width of one component induces a near unit change in the other. Therefore, the relative fluctuations are in the ratio of the mean values of these parameters ($\sim100:1$). Similarly, the parameters associated with amplitudes, the optical depth $\alpha$ and the optical split ratio $C$, have similar absolute deviations from their mean values. However, the mean value of the depth is $\sim1/20$ of the background level due to the chosen cell temperature. The flicker-noise prevent the uncertainty in the fitted parameters decreases with ensemble averaging after a few scans. 


\end{document}